\documentclass[twocolumn,prl,showpacs,amsfonts,amsmath,assymb,eufrak]{revtex4}
\def\eq#1\en{\begin{equation}#1\end{equation}}  
\def\eqa#1\ena{\begin{align}#1\end{align}}
\def\eqg#1\eng{\begin{gather}#1\end{gather}}
\newcommand{\lb}[1]{\label{e:#1}}
\newcommand{\rlb}[1]{\eqref{e:#1}} 
\newcommand{\nl}{\notag\\}

\newcommand{\sbkt}[1]{\langle#1\rangle}

\newcommand{\sumtwo}[2]%
{\mathop{\sum_{#1}}_{#2}}
\newcommand{\sumthree}[3]%
{\mathop{\mathop{\sum_{#1}}_{#2}}_{#3}}
\newcommand{\sumfour}[4]%
{\mathop{\mathop{\mathop{\sum_{#1}}_{#2}}_{#3}}_{#4}} 
\newcommand{\prodtwo}[2]%
{\mathop{\prod_{#1}}_{#2}}
\newcommand{\mintwo}[2]%
{\mathop{\min_{#1}}_{#2}}
\newcommand{\maxtwo}[2]%
{\mathop{\max_{#1}}_{#2}}
\newcommand{\maxthree}[3]%
{\mathop{\mathop{\max_{#1}}_{#2}}_{#3}}
\newcommand{\limtwo}[2]%
{\mathop{\lim_{#1}}_{#2}}
\newcommand{\suptwo}[2]%
{\mathop{\sup_{#1}}_{#2}}
\newcommand{\supthree}[3]%
{\mathop{\mathop{\sup_{#1}}_{#2}}_{#3}}
\newcommand{\supfour}[4]%
{\mathop{\mathop{\mathop{\sup_{#1}}_{#2}}_{#3}}_{#4}} 
\newcommand{\inftwo}[2]%
{\mathop{\inf_{#1}}_{#2}}
\newcommand{\infthree}[3]%
{\mathop{\mathop{\inf_{#1}}_{#2}}_{#3}}
\newcommand{\inffour}[4]%
{\mathop{\mathop{\mathop{\inf_{#1}}_{#2}}_{#3}}_{#4}} 

\newcommand\calH{{\cal H}}

\newcommand{\mrx}{\mathrm{x}}
\newcommand{\mry}{\mathrm{y}}
\newcommand{\mrz}{\mathrm{z}}

\newcommand{\bsr}{\boldsymbol{r}}
\newcommand{\bss}{\boldsymbol{s}}

\newcommand{\bbR}{\mathbb{R}}

\newcommand{\pmz}{\{+,-,0\}}
\newcommand{\hpsi}{\hat{\psi}}
\newcommand{\hpsid}{\hat{\psi}^\dagger}
\newcommand{\hrho}{\hat{\rho}}
\newcommand{\dr}{d^3\bsr}

\newcommand{\Hs}{\hat{H}_0}
\newcommand{\OH}{\hat{O}_\mathrm{H}}

\newcommand{\hV}{\hat{V}}

\newcommand{\ad}{\hat{a}^\dagger}

\newcommand{\PGS}{\Phi_\mathrm{GS}}
\newcommand{\PGSH}{\PGS^\mathrm{H}}

\newcommand{\Pv}{\Phi_\mathrm{vac}}
\newcommand{\Pf}{\Phi^\mathrm{free}_0}

\newcommand{\hS}{\hat{S}}

\newcommand{\Nz}{\hat{N}_0}
\newcommand{\hbS}{\hat{\mathbf{S}}}
\newcommand{\Ham}{\hat{H}}

\newcommand{\Pp}{\Phi^\mathrm{free}_\mathrm{pol}}
\newcommand{\PAF}{\Phi^\mathrm{free}_\mathrm{AF}}
\newcommand{\proj}{\hat{P}_0}
\newcommand{\midskip}{\vspace{3pt}}

\begin{document}
\title{``Polar'' and ``antiferromagnetic'' order in $f=1$ many-boson systems}

\author{Hal Tasaki}
\affiliation{Department of Physics, Gakushuin University, 
Mejiro, Toshima-ku, Tokyo 171-8588, Japan}

\date{\today}

\begin{abstract}
In a system of  interacting $f=1$ bosons (in the subspace where the total spin in the z-direction is vanishing), we prove inequalities for the ground state expectation value of the density of spin-0 bosons.
The inequalities imply that the ground state possesses ``polar'' or ``antiferromagnetic'' order when the quadratic Zeeman term $q$ is large enough.
In the low density limit, the inequalities establish the existence of a sharp transition at $q=0$ when $q$ is varied.
\end{abstract}


\maketitle

Recent progress in cold atom experiments has led to a renewed interest in the low energy properties of interacting many-boson systems.
Especially the systems of spinful bosons \cite{Stamper-Kurn,Stenger,BarrettSauerChapman,Chang}, in which hyperfine spin degrees of freedom couple to many-body quantum physics of bosons, are expected to show a variety of nontrivial quantum magnetic phenomena.
See \cite{KawaguchiUeda} for a recent review.
It is a fascinating challenge in many-body physics to give a firm (hopefully rigorous) theoretical understanding of these phenomena \cite{LiebSeiringerSolovejYngvason}.
The present work is a first step in such a direction.

It has been argued since the early days \cite{OhmiMachida,Ho} that $f=1$ Bose-Einstein condensates with antiferromagnetic interaction have peculiar magnetic phases called ``polar'' and ``antiferromagnetic'', in which global spin rotational symmetry is spontaneously broken \cite{Stenger}.
This symmetry breaking is of a ``nematic'' type (in the sense that the symmetry breaking in, e.g., the positive z-direction is the same as that in the negative z-direction).
Interestingly the existence of nematic long-range order has been established rigorously in a class of $S=1$ quantum spin systems \cite{TanakaTanakaIdogaki} which can be regarded as limits of the $f=1$ Bose-Hubbard model \cite{Imambekov,Snoek}.
The nematic phases in Bose-Einstein condensates and in the spin systems should be related with each other.

Magnetic properties of interacting boson system have been studied mainly by mean field theories, and few rigorous results are known.
Here we present rigorous inequalities for the ground states (in the subspace where the total spin in the z-direction is vanishing) of fully interacting spinful boson systems which imply the existence of ``polar'' or ``antiferromagnetic'' order when the quadratic Zeeman term is large enough.
In particular we can show the existence of a sharp transition in the low density limit.
We stress that this is not only one of the first rigorous results in  spinful many-boson systems, but also one of (not so many) rigorous results in many-body physics that apply to realistic models and produce experimentally verifiable concrete results.

We also discuss weak coupling theory, and place the present problem into the context of the standard picture of long-range order and symmetry breaking.
We believe that this observation dissolves the apparent (and well-known) discrepancy between the predictions of the exact weak coupling theory \cite{LawPuBigelow,KoahiUeda,HoYip} and of the mean-field theory \cite{KawaguchiUeda,OhmiMachida,Ho}.

We hope that the present study can become a starting point of further mathematical physical researches on quantum magnetic phenomena in interacting boson systems.

\midskip
\paragraph*{Definitions:}
We study a system of $N$ bosons with (hyperfine) spin $f=1$ in $\bbR^3$ (or in a finite region of $\bbR^3$), where $N$ is even.
Extension to bosons with higher spins is trivial.
As usual we denote by $\hpsi_\sigma(\bsr)$ and $\hpsid_\sigma(\bsr)$ the annihilation and the creation operators, respectively, of a boson at $\bsr\in\bbR^3$ with (the z-component of) spin $\sigma\in\pmz$.
These operators satisfy the commutation relations $[\hpsi_\sigma(\bsr),\hpsid_{\sigma'}(\bsr')]=\delta_{\sigma,\sigma'}\delta(\bsr-\bsr')$ for any $\sigma,\sigma'\in\pmz$ and $\bsr,\bsr'\in\bbR^3$.
More rigorously, for any single-particle state (wave function) $\varphi=(\varphi(\bsr))_{\bsr\in\bbR^3}$, the operator $\hpsid_\sigma[\varphi]:=\int\dr\,\varphi(\bsr)\,\hpsid_\sigma(\bsr)$ creates a boson in the state $\varphi$ with spin $\sigma$.
We also let $\hrho_\sigma(\bsr):=\hpsid_\sigma(\bsr)\hpsi_\sigma(\bsr)$ be the density of the spin $\sigma$ bosons, and $\hrho(\bsr)=\sum_{\sigma\in\pmz}\hrho_\sigma(\bsr)$ the density.

We consider the Hilbert space $\calH_0$, which is spanned by states of the form $\{\prod_{j=1}^N\hpsid_{\sigma_j}[\varphi_j]\}\,\Pv$, where $\Pv$ is the state with no bosons in the trap, $\varphi_j$ are arbitrary single-particle states (in a suitable space), and $\sigma_j\in\pmz$ are arbitrary except for the condition $\sum_{j=1}^N\sigma_j=0$.
Note that we have restricted ourselves to the space where the z-component of the total spin is vanishing.
This choice is realistic since our Hamiltonian \rlb{H} conserves  $\sum_{j=1}^N\sigma_j$.
Experimentalists can prepare an initial state in $\calH_0$, which should evolve into a ground state in $\calH_0$.

Let us write the full Hamiltonian as
\eq
\Ham=\Hs+\hV-q\Nz.
\lb{H}
\en
The Hamiltonian of the non-interacting system is 
\eq
\Hs:=\int\dr\Bigl\{
-\frac{\hbar^2}{2m}
\hspace{-3pt}\sum_{\sigma\in\pmz}
\hspace{-3pt}
\hpsid_\sigma(\bsr)\Delta\hpsi_\sigma(\bsr)
+U(\bsr)\,\hrho(\bsr)
\Bigr\},
\en
where $U(\bsr)$ is a suitable single-particle potential which describes the trap.

We can consider an essentially arbitrary interaction $\hV$ which is nonnegative and leaves the space $\calH_0$ invariant.
To be concrete, however, let us take the most standard low energy effective interaction of dilute $f=1$ bosons $\hV=\hV_0+\hV_2$, where
\eq
\hV_0:=\frac{1}{2}\int\dr\,\dr'\,v_0(\bsr-\bsr')
\hat{A}^\dagger_{0,0}(\bsr,\bsr')\hat{A}_{0,0}(\bsr,\bsr')
\lb{F=0}
\en
and
\eq
\hV_2:=\frac{1}{2}\int\dr\,\dr'\,v_2(\bsr-\bsr')
\sum_{m=-2}^2\hat{A}^\dagger_{2,m}(\bsr,\bsr')\hat{A}_{2,m}(\bsr,\bsr')
\lb{F=2}
\en
represent the interactions between two bosons with total spin 0 and 2, respectively.
See, e.g., \cite{KawaguchiUeda}. 
Here $v_0(\bss)$ and $v_2(\bss)$ are suitable nonnegative potentials, and  $\hat{A}_{F,m}(\bsr,\bsr'):=\sum_{\sigma,\sigma'}\langle F,m|\sigma,\sigma'\rangle\,\hpsi_\sigma(\bsr)\hpsi_{\sigma'}(\bsr')$ (where $\langle F,m|\sigma,\sigma'\rangle$ denotes the Clebsh-Goldan coefficients) is the annihilation operator for the state in which the total spin and its z-component of two bosons (at $\bsr$ and $\bsr'$) are $F$ and $m$, respectively \cite{A}.

In the literature one often considers the local interactions with $v_F(\bss)=g_F\delta(\bss)$ where $F=0,2$.
Here, to avoid mathematical complications, we do not treat the delta function interactions.
But we shall discuss the limit of short-range interaction after presenting the main inequalities in Theorem~2.

Finally $-q\Nz$ in \rlb{H} is the quadratic Zeeman term generated by a uniform external magnetic field, where $q\in\bbR$ and $\Nz=\int\dr\,\hrho_0(\bsr)$ is the total number of spin-0 bosons.
The standard Zeeman term is omitted since it has no effects within the space $\calH_0$ as long as the magnetic field is uniform.

\midskip
\paragraph*{Long-range order and spontaneous symmetry breaking in the weak coupling limit:}
To put the problem into the context of long-range order and spontaneous symmetry breaking in many-body systems, we shall discuss (almost trivial) weak coupling limit in some detail.

Consider the system whose Hamiltonian consists only of the non-interacting part $\Hs$.
Since $\Hs$ does not flip the spin of each particle, the model has highly degenerate ground states within $\calH_0$ \cite{EisenbergLieb}.

Suppose that one adds to $\Hs$ a small spin-dependent interaction $\hV$ which favors two particles to have vanishing total spin.
This should be the case if  $v_0(\bss)\le v_2(\bss)$.
Then it is expected (and can be proved in related models \cite{KT}) that the degeneracy is immediately lifted and the ground state in $\calH_0$ is unique and has vanishing total spin.

Let us consider this unique ground state in the limit where $\hV$ vanishes, and denote it as $\Pf$.
Since we are now dealing with a non-interacting system, the $N$-body ground state is written in terms of the creation operators $\ad_{\sigma}:=\int\dr\,\varphi_0(\bsr)\hpsid_\sigma(\bsr)$, where $\varphi_0$ is the unique normalized ground state of the single-particle Schr\"odinger equation $-\{\hbar^2/(2m)\}\Delta\varphi(\bsr)+U(\bsr)\varphi(\bsr)=\epsilon\,\varphi(\bsr)$.
It is also known that the information that the ground state has vanishing total spin is enough to uniquely determine $\Pf$ as
\eq
\Pf=\sumtwo{n=0}{(n:\text{even})}^{N}
\frac{C\,(-2)^{-n/2}}{\frac{n}{2}!\frac{N-n}{2}!}(\ad_0)^n(\ad_+\ad_-)^{(N-n)/2}\,\Pv,
\lb{freeGS}
\en
where $C$ is a normalization constant \cite{LawPuBigelow,KoahiUeda,HoYip}.
This in particular implies that the probability $\text{Prob}[\Nz=n]$ that the number $\Nz$ of spin-0 bosons is equal to $n$ behaves as
\eq
\text{Prob}[\Nz=n]\propto\frac{n!}{2^n\{(n/2)!\}^2}
\approx\frac{1}{\sqrt{n}}.
\lb{P0}
\en

Let us define the ``order parameter'' by $\hat{O}:=\Nz-(N/3)$.
The rotation invariance of $\Pf$ implies $\sbkt{\Pf,\hat{O}\,\Pf}=0$ (see Theorem~1 below), which also follows from \rlb{P0}.
The peculiar power law behavior of \rlb{P0} manifests itself in the fluctuation of $\hat{O}$ as $\sbkt{\Pf,\hat{O}^2\,\Pf}\simeq(4/45)N^2$, which becomes enormously large when $N$ is large.

Such a behavior of order parameter is indeed typical in a many-body state which exhibits  long-range order but no spontaneous symmetry breaking \cite{KomaTasaki}.
As a well-understood typical example, consider the $S=1/2$ antiferromagnetic Heisenberg model on the $L\times L\times L$ cubic lattice with the Hamiltonian $\Ham_\mathrm{H}=\sum_{\langle{j,j'\rangle}}\hat{\boldsymbol{S}}_j\cdot\hat{\boldsymbol{S}}_{j'}$ (where the sum is over all the neighboring sites), and define the Ne\'el order parameter (or the staggered magnetization) by $\OH:=\sum_j(-1)^{j_1+j_2+j_3}\hat{S}_j^{(\mathrm{z})}$ (where we labeled the sites as $j=(j_1,j_2,j_3)$).
Although the present model exhibits Ne\'el order at low temperatures and in the ground state, the ground state $\PGSH$ is unique for any finite $L$.
The uniqueness implies $\sbkt{\PGSH,\OH\PGSH}=0$, giving no indication of the Ne\'el order.
The order manifests itself in the large fluctuation of $\OH$ (or, equivalently, antiferromagnetic long-range order) as $\sbkt{\PGSH,(\OH)^2\PGSH}\ge(\text{const.})(L^3)^2$ \cite{DysonLiebSimon,JordaoNeves,KennedyLiebShastry}.
The ground state $\PGSH$ exhibits long-range order but no symmetry breaking.
To get a ground state which exhibits both long-range order and  symmetry breaking, one introduces the fictitious ``staggered magnetic field'' $h$.
For $h>0$ let $\PGS^{\mathrm{H},h}$ be the ground state of $\Ham_\mathrm{H}-h\OH$.
Then it is proved \cite{Kaplan} that the spontaneous staggered magnetization satisfies
\eq
m_\mathrm{s}:=\lim_{h\downarrow0}\lim_{L\uparrow\infty}\frac{1}{L^3}\sbkt{\PGS^{\mathrm{H},h},\OH\PGS^{\mathrm{H},h}}>0.
\lb{ms}
\en
One can say that an infinitesimally small symmetry breaking field $h$  (in the infinite system) has triggered the symmetry breaking.
Although a large but finite system does not exhibit symmetry breaking, it is known to be extremely sensitive to staggered magnetic field $h$, easily yielding a state with large staggered magnetic field.

The situation with our ground state $\Pf$ is quite similar \cite{BarnettSauSarma}.
Here the quadratic Zeeman term $-q\Nz$ precisely plays the role of the symmetry breaking field.
Indeed it is easily seen (for the Hamiltonian $\Hs-q\Nz$) that the ground state for any $q>0$ is the ``polar'' state
\eq
\Pp:=\frac{1}{\sqrt{N!}}(\ad_0)^N\Pv,
\lb{polar}
\en
which has $\sbkt{\Pp,\Nz\Pp}=N$, and the ground state for any $q<0$ is the ``antiferromagnetic'' state \cite{AFnote}
\eq
\PAF:=\frac{1}{(N/2)!}(\ad_+\ad_-)^{N/2}\Pv,
\lb{AF}
\en
which has $\sbkt{\PAF,\Nz\PAF}=0$.
We can say that the ground state \rlb{freeGS} (in which $\hat{O}$ fluctuates widely) is extremely sensitive to the applied field $q$;
infinitesimally small $q$ leads to stable states \rlb{polar} or \rlb{AF}, which break the spin rotation symmetry explicitly.
This is the basic reason that the mean-field theory predicts the polar or antiferromagnetic ground states even for $q=0$ \cite{KawaguchiUeda,OhmiMachida,Ho} in an apparent disagreement with the exact result \rlb{freeGS} for weak coupling \cite{LawPuBigelow,KoahiUeda,HoYip}.

We note in passing that the state \rlb{AF}, which in the form of fragmented Bose-Einstein condensate \cite{MuellerHoUedaBaym}, is also rewritten in a ``mean-field form''
\eq
\PAF=(\text{const.})\,\proj(\ad_+-e^{2i\theta}\ad_-)^N\Pv,
\lb{AF2}
\en
where $\proj$ is the projection onto $\calH_0$.
Note that \rlb{AF2} is independent of $\theta\in\bbR$.
Here $(\ad_+-e^{2i\theta}\ad_-)/\sqrt{2}$ creates the eigenstate of $\cos\theta\,\hS^{(\mrx)}+\sin\theta\,\hS^{(\mry)}$ (i.e., the spin component in a direction orthogonal to the z-axis) with the eigenvalue 0.
Since a suitable rotation in spin space brings $\ad_0$ into $(\ad_+-e^{2i\theta}\ad_-)/\sqrt{2}$, one might be tempted to conclude that the polar state \rlb{polar} and the antiferromagnetic state \rlb{AF}, \rlb{AF2} are essentially the same state.
But this is not the case since there is a projection in \rlb{AF2}; the two states are distinct \cite{MFnote}.

We stress that the polar state \rlb{polar} and the antiferromagnetic state \rlb{AF} appear as a consequence of a breakdown of the rotational symmetry.
The symmetry breaking is triggered by the applied field $-q\Nz$, which breaks the symmetry in the z-direction.
This direction is arbitrary.
If the symmetry is broken in the x-direction, for example, the polar and the antiferromagnetic states become $(\text{const.})(\ad_+-\ad_-)^N\Pv$ and $(\text{const.})\{(\ad_++\sqrt{2}\,\ad_0+\ad_-)(\ad_+-\sqrt{2}\,\ad_0+\ad_-)\}^{N/2}\Pv$, respectively.
We also note that the present symmetry breaking does not distinguish between, e.g., the positive and the negative z-direction.
In this sense the present rotational symmetry breaking is {\em nematic}\/.

\midskip
\paragraph*{Main results:}
Let us turn to interacting models and discuss our theorems.
Although we believe that the ground state of the present model exhibits Bose-Einstein condensation, we do not make use of that property here.
We shall focus on the ground state expectation value of $\hrho_0:=\int\dr\,\hrho_0(\bsr)=\Nz/N$, the ratio of the spin-0 bosons.
We first make the following elementary observation.

{\em Theorem 1:}\/ Let $q=0$, and suppose that the ground state $\PGS$ of the full Hamiltonian \rlb{H} is unique and has vanishing total spin.
Then we have $\sbkt{\PGS,\hrho_0\PGS}=1/3$.

The unique ground state with vanishing total spin is expected for ``antiferromagnetic'' interactions characterized by $v_0(\bss)\le v_2(\bss)$.
See \cite{KT}.

As before let $\varphi_0(\bsr)$ be the unique normalized ground state of the single-particle Schr\"odinger equation corresponding to \rlb{H}.
Let us define the effective volume $V$ as the smallest constant which satisfies $|\varphi_0(\bsr)|^2\le1/V$ for any $\bsr$.
We also let $g_F:=\int d^3\bss\, v_F(\bss)\ge0$ for $F=0,2$. 

{\em Theorem 2:}\/ 
Let $\PGS$ be any ground state (in $\calH_0$) of the full Hamiltonian \rlb{H}.
For any $q>0$, one has
\eq
\sbkt{\PGS,\hrho_0\PGS}\ge1-\frac{g_0+2g_2}{6q}\frac{N}{V},
\lb{21}
\en
and, for any $q<0$, one has
\eq
\sbkt{\PGS,\hrho_0\PGS}\le\frac{2g_0+g_2}{12|q|}\frac{N}{V}.
\lb{22}
\en

The inequalities \rlb{21}, \rlb{22} imply that the ground state of \rlb{H} exhibits polar or antiferromagnetic order in the sense that $\sbkt{\PGS,\hrho_0\PGS}$ is close to 1 or 0, respectively,  when $|q|\gg gN/V$ (assuming $g\simeq g_0\simeq g_2$) \cite{Nanote}.
Note that one cannot expect a sharp transition as in the free model discussed in the previous part.
Since the ground state (within $\calH_0$) for a general nonvanishing $\hV$ is expected to be unique for any $q\in\bbR$, the expectation value $\sbkt{\PGS,\Nz\PGS}$ should vary continuously with $q$ when $N$ is finite.
This is a manifestation of the general principle that one generally does not observe symmetry breaking in a finite system.

To see a sharp transition (as $q$ is varied), one should take a certain limit.
The most challenging and interesting setup is to consider a model with suitable antiferromagnetic interaction and take the infinite volume limit (with a fixed density $\rho$) as in the case of the Heisenberg model \rlb{ms}.
But this problem seems to be formidably difficult from the current technique of many-body systems.

What one can conclude from the inequalities \rlb{21}, \rlb{22} is
\eq
\lim_{(N/V)\downarrow0}\sbkt{\PGS,\hrho_0\PGS}=
\begin{cases}
1,&q>0,\\0,&q<0,
\end{cases}
\en
which establishes the existence of a sharp transition in the low density limit.
Note that this result is valid no matter how strong the interaction is.
This provides a partial rigorous justification of the prediction by the mean-field theory \cite{KawaguchiUeda}.

Since the inequalities in Theorem~2 depend on the interaction only through $g_0$ and $g_2$, it is easy to consider the limit of short range interaction.
One starts with  potentials $v_0(\bss)$, $v_2(\bss)$ which are sufficiently ``nice'' so that the ground state $\PGS$ is well defined.
After stating the inequalities \rlb{21}, \rlb{22} for $\PGS$, one may formally take the limit $v_F(\bsr)\to g_F\,\delta(\bss)$ where $F=0,2$.
The inequalities \rlb{21}, \rlb{22} remain valid as they are (provided that the limits of the left-hand sides are well defined).

\midskip
\paragraph*{Proof:}
To prove Theorem~1, we note that the conditions of the theorem implies that the ground state $\PGS$ is invariant under rotation in spin space, and any rotationally invariant state $\Phi$ satisfies $\sbkt{\Phi,\Nz\Phi}=N/3$.
One way to see the latter fact is to rewrite $\Phi$ in the language of distinguishable identical particles (by working out the symmetrization explicitly).
Let $\hbS_j$ be the spin operator of the $j$-th particle.
Since $(\hbS_j)^2=2$, the rotational symmetry implies $\sbkt{\Phi,(\hS^{(\mrz)}_j)^2\Phi}=2/3$, which implies that the probability that the $j$-th particle has spin-0 is $1/3$.

Theorems~2 is based on the following simple variational argument.
For any $\Psi\in\calH_0$, one has $\sbkt{\PGS,H\PGS}\le\sbkt{\Psi,H\Psi}$.
By using \rlb{H}, we rewrite this as
\eqa
q&\sbkt{\PGS,\Nz\PGS}
\nl&\ge q\sbkt{\Psi,\Nz\Psi}+E_0-\sbkt{\Psi,\Hs\Psi}
-\sbkt{\Psi,\hV\Psi},
\lb{main2}
\ena
where we noted  $\sbkt{\PGS,\hV\PGS}\ge0$ and $\sbkt{\PGS,\Hs\PGS}\ge E_0$, where $E_0$ is the ground state energy of $\Hs$.

For $q>0$, we choose the variational state $\Psi$ as the polar state $\Pp$ of \rlb{polar}.
Then one readily sees that $\sbkt{\Psi,\Nz\Psi}=N$ and $\sbkt{\Psi,\Hs\Psi}=E_0$.
The remaining $\sbkt{\Psi,\hV\Psi}$ can be evaluated explicitly as follows.
Observe first that
\eq
\hat{A}^\dagger_{0,0}(\bsr,\bsr')\hat{A}_{0,0}(\bsr,\bsr')
=\frac{1}{3}\hpsid_0(\bsr')\hpsid_0(\bsr)\hpsi_0(\bsr)\hpsi_0(\bsr')+\cdots,
\en
where the omitted terms involve spin $+$ and $-$, and has no contributions to $\sbkt{\Pp,\hV_0\Pp}$.
By noting that
\eqa
\Psi'&:=\hpsi_0(\bsr)\hpsi_0(\bsr')\Pp
\nl&
=N(N-1)\varphi_0(\bsr)\varphi_0(\bsr')\frac{(\ad_0)^{N-2}}{\sqrt{N!}}\Pv,
\ena
we find
\eqa
\sbkt{\Pp&,\hV_0\Pp}
\nl&=\frac{1}{6}\int\dr\dr'
v_0(\bsr-\bsr')\sbkt{\Psi',\Psi'}
\nl&
=\frac{N(N-1)}{6}\int\dr\dr'
v_0(\bsr-\bsr')|\varphi_0(\bsr)|^2|\varphi_0(\bsr')|^2
\nl&
\le\frac{g_0N^2}{6V},
\ena
where we used $|\varphi_0(\bsr')|^2\le1/V$ to get the final line.
By evaluating $\sbkt{\Pp,\hV_2\Pp}$ in a similar manner, we get the desired inequality \rlb{21}.

For $q<0$, we choose the variational state $\Psi$ as the antiferromagnetic state $\PAF$ of \rlb{AF}, and proceed in the similar manner to get \rlb{22}.

\midskip
\paragraph*{Discussions:}
We have proved the inequalities which shows that an interacting $f=1$ boson system exhibits polar or ``antiferromagnetic'' order when sufficiently large quadratic Zeeman term is present.
Although the conclusion may not be surprising, we stress that it has been established rigorously in a system with arbitrarily strong interaction.
We hope that the present study becomes a starting point of further mathematical works on quantum magnetic phenomena in interacting boson systems. 

We have also placed the present problem of nematic order in boson systems into the larger context of long-range order and symmetry breaking in quantum many-body systems.
That discussion as well as the mean-field analysis \cite{KawaguchiUeda,OhmiMachida,Ho} suggests that the transitions to the polar or antiferromagnetic states are more drastic when the interaction has an antiferromagnetic nature \cite{ferro}.
This picture is consistent with the fact that the existence of nematic long-range order is proved in  $S=1$ quantum spin systems which can be regarded as limits of the $f=1$ Bose-Hubbard model with antiferromagnetic interaction  \cite{TanakaTanakaIdogaki,Imambekov,Snoek}.
Unfortunately our inequalities do not make any essential distinctions between antiferromagnetic and ferromagnetic interactions.
It is an interesting theoretical challenge to go beyond the present naive variational argument to see essential roles played by antiferromagnetic interactions.

\midskip
{\small 
I wish to thank Tohru Koma for valuable discussions and for pointing me out a flaw in the earlier version of the paper.
I also thank  
Takuya Hirano,
Hosho Katsura,
Yuki Kawaguchi,
Masahiro Takahashi,
Tatsuyoshi Tanabe,
Akinori Tanaka,
Yuta Toga,
and
Masahito Ueda
for valuable discussions.
}



\begin{thebibliography}{10}
\bibitem{A}
Explicit forms are 
$\hat{A}_{2,2}=\hpsi_+\hpsi'_+$,
$\hat{A}_{2,1}=\{\hpsi_+\hpsi'_0+\hpsi_0\hpsi'_+\}/\sqrt{2}$,
$\hat{A}_{2,0}=\{\hpsi_+\hpsi'_-+2\hpsi_0\hpsi'_0+\hpsi_-\hpsi'_+\}/\sqrt{6}$,
$\hat{A}_{2,-1}=\{\hpsi_-\hpsi'_0+\hpsi_0\hpsi'_-\}/\sqrt{2}$,
$\hat{A}_{2,-2}=\hpsi_+\hpsi'_+$,
and
$\hat{A}_{0,0}=\{\hpsi_+\hpsi'_--\hpsi_0\hpsi'_0+\hpsi_-\hpsi'_+\}/\sqrt{3}$,
where $\hpsi_\sigma$ and $\hpsi'_\sigma$ are shorthands for $\hpsi_\sigma(\bsr)$ and $\hpsi_\sigma(\bsr')$, respectively.
\bibitem{AFnote}
This state has very little in common with ``antiferromagnetic states'' (in the sense of states with antiferromagnetic correlation) studied in spin systems.
\bibitem{MFnote}
In a mean-field analysis (see, e.g., \cite{KawaguchiUeda}), one usually omits the projection $\proj$.
This is justified since $\proj$ does not affect bulk properties in the limit $N\uparrow\infty$.
We nevertheless believe it advantageous to keep $\proj$ (even in heuristic discussions) since it makes it clear, e.g.,  that $\Pp$ and $\PAF$ are distinct, or that \rlb{AF2} is independent of $\theta$.
\bibitem{Nanote}
For ${}^{23}$Na, one has $g\sim5\times10^{-52}\rm\ J/m^3$.
By setting $N/V\sim10^{20}\rm/m^3$, we get $(gN/V)/h\sim1\times10^2\rm\ Hz$, which is an easily realizable value of $|q|$.
\bibitem{ferro}
The polar state is observed experimentally in a ferromagnetic system with a suitable quadratic Zeeman field \cite{Chang}.
\end{thebibliography}
\end{document}